\documentclass[final,times,twocolumn]{elsarticle}
\usepackage{amssymb}
\usepackage{hyperref}
\usepackage{amsmath}
\usepackage{url}
\usepackage{marginnote}
\usepackage{xcolor}
\biboptions{sort&compress}

\journal{Physics Letters B}

\allowdisplaybreaks[2]

\newcommand{\df}{\mathrm{d}}

\newcommand{\nn}{\nonumber}

\begin{document}

\begin{frontmatter}

\title{$\nu$-point energy correletors with \textsc{FastEEC}: small-$x$ physics from LHC jets}

\author[Nikhef]{Ankita Budhraja}

\author[MIT]{Hao Chen}

\author[Nikhef,UvA]{Wouter J.~Waalewijn}

\affiliation[Nikhef]{organization={Nikhef, Theory Group},
              addressline={Science Park 105}, 
              postcode={1098 XG}, 
              city={Amsterdam} , 
              country={The Netherlands}}

\affiliation[MIT]{organization={Center for Theoretical Physics, Massachusetts Institute of Technology},
		addressline={77 Massachusetts Avenue},
		city={Cambridge},
		postcode={MA 02139},
		country={U.S.A}}
            
\affiliation[UvA]{organization={Institute of Physics and Delta Institute for Theoretical Physics, University of Amsterdam},
            addressline={Science Park 904}, 
            postcode={1098 XH}, 
            city={Amsterdam},
            country={The Netherlands}}

\tnotetext[preprint]{Preprint number: MIT--CTP 5763}

\begin{abstract}
In recent years, energy correlators have emerged as a powerful tool for studying jet substructure, with promising applications such as probing the hadronization transition, analyzing the quark-gluon plasma, and improving the precision of top quark mass measurements. The projected $N$-point correlator measures correlations between $N$ final-state particles by tracking the largest separation between them, showing a scaling behavior related to DGLAP splitting functions. These correlators can be analytically continued in $N$, commonly referred to as $\nu$-correlators, allowing access to non-integer moments of the splitting functions. Of particular interest is the $\nu \to 0$ limit, where the small momentum fraction behavior of the splitting functions requires resummation.
Originally, the computational complexity of evaluating $\nu$-correlators for $M$ particles scaled as $2^{2M}$, making it impractical for real-world analyses. However, by using recursion, we reduce this to $M 2^M$, and through the \textsc{FastEEC} method of dynamically resolving subjets, $M$ is replaced by the number of subjets. This breakthrough enables, for the first time, the computation of $\nu$-correlators for LHC data. In practice, limiting the number of subjets to 16 is sufficient to achieve percent-level precision, which we validate using known integer-$\nu$ results and convergence tests for non-integer $\nu$.
We have implemented this in an update to \textsc{FastEEC} and conducted an initial study of power-law scaling in the perturbative regime as a function of $\nu$, using CMS Open Data on jets. The results agree with DGLAP evolution, except at small $\nu$, where the anomalous dimension saturates to a value that matches the BFKL anomalous dimension. 
\end{abstract}


\end{frontmatter}

\section{Introduction}
\label{sec:introduction}

The Large Hadron Collider (LHC) is an ideal machine to study Quantum Chromodynamics (QCD) over a wide range, from asymptotically-free quarks and gluons to hadron confinement. The concept of a QCD jet, which is a collimated collection of hadrons produced in high-energy collisions, provides a crucial stage for understanding perturbative parton dynamics as well as non-perturbative hadronization. 
While the community has converged on the anti-$k_T$ algorithm~\cite{Cacciari:2008gp} to describe jets, many jet-based observables have been proposed to probe the internal properties of jets (see ~\cite{Larkoski:2017jix, Kogler:2018hem, Marzani:2019hun} for reviews). 

Energy correlators were first proposed in~\cite{Basham:1977iq, Basham:1978bw, Basham:1978zq, Basham:1979gh} and used at the Large Electron-Positron collider  (LEP)~\cite{OPAL:1991uui}, where the primary focus was on the distribution of jets. 
In recent years, energy correlators have found fruitful applications for characterizing the energy flow patterns inside jets, due to their natural separation of scales and suppression of soft radiation effects. 
The original concept was generalized in~\cite{Hofman:2008ar,Chen:2020vvp}, which is argued to be a complete family of infrared (IR) safe observables and well-suited for phenomenology in jet substructure~\cite{Chen:2020vvp}.

From a theoretical perspective, energy correlators inside jets can be described by collinear factorization~\cite{Dixon:2019uzg,Chen:2020vvp}, or equivalently the light-ray Operator Product Expansion (OPE)~\cite{Hofman:2008ar,Kologlu:2019mfz,Chang:2020qpj,Chen:2023zzh}, exhibiting an interesting (approximate) scaling behavior. At the LHC, the excellent performance of modern detectors and large jet ensembles bring these theoretical considerations to life in measurements. Initiated in~\cite{Komiske:2022enw} using CMS Open Data, energy correlators have been measured at ALICE~\cite{Mazzilli:2024ots}, CMS~\cite{CMS:2024mlf} and STAR~\cite{Tamis:2023guc}. In particular, it currently gives the most accurate $\alpha_s$ extraction from jet substructure~\cite{CMS:2024mlf}. 
Other important phenomenological applications include the determination of the top quark mass~\cite{Holguin:2022epo, Holguin:2023bjf, Holguin:2024tkz}, studying the hadronization transition in QCD~\cite{Komiske:2022enw, Lee:2024esz}, understanding the dead-cone effect~\cite{Craft:2022kdo}, probing the emergent scales in the presence of a thermal medium~\cite{Andres:2022ovj, Andres:2023xwr, Barata:2023zqg, Barata:2023bhh, Andres:2024ksi, Singh:2024vwb}, as well as unveiling  gluon saturation in electron-ion collisions~\cite{Liu:2023aqb} using a nucleon energy correlator~\cite{Liu:2022wop}. 
Performing measurements of energy correlators on tracks benefits from the superior angular resolution with which tracks can be measured. The corresponding precision calculations have been developed in Refs.~\cite{Chen:2020vvp, Li:2021zcf, Chen:2022muj, Chen:2022pdu, Jaarsma:2023ell}, using the track function formalism~\cite{Chang:2013rca, Chang:2013iba}.
There are also interesting formal theory developments for energy correlators in the collinear limit~\cite{Chen:2019bpb, Chen:2020adz, Chen:2021gdk, Schindler:2023cww, Gao:2023ivm, Chen:2023zlx, Chicherin:2024ifn, Chen:2024nyc, Chang:2022ryc,Chen:2024iuv}.

In this letter, we focus on the projected $\nu$-point energy correlators ($\nu$-correlators), proposed in Ref.~\cite{Chen:2020vvp} as an analytic continuation of the projected (integer) $N$-point correlators, defined as
\begin{align}
\frac{d\sigma^{[N]}}{d R_L} &=\sum_X \int d\sigma_X \sum_{i_1,i_2,\dots, i_N\in X} z_{i_1} z_{i_2} \dots z_{i_N} \nonumber\\
&\times \delta(R_L-\mathrm{max}\{\Delta R_{i_1, i_2}, \Delta R_{i_1, i_3}\dots \Delta R_{i_{N-1}, i_N}\})\,. \label{eq:PENC_def}
\end{align}
Here $d\sigma_X$ is the differential cross section of producing the final state $X$, $z_i=p_{T,i}/p_{T,\text{jet}}$ is the momentum fraction of particle $i$. $\Delta R_{ij}$ is the angular variable between particle $i$ and $j$, which can be approximated in the collinear limit, in terms of rapidity-azimuthal variable $(y,\phi)$, as
\begin{equation}
\Delta R_{ij} = \sqrt{(\Delta y_{ij})^2+ (\Delta \phi_{ij})^2}\,.
\end{equation}
The delta function in Eq.~\eqref{eq:PENC_def} projects the observable from $\frac{1}{2}N(N-1)$ angular variables to the largest angle, denoted as $R_L$. In the following, we also consider the equivalent cumulative version
\begin{equation}
\Sigma^{[N]}(R_L) = \int_0^{R_L}\! \df R_L'\, \frac{d\sigma^{[N]}}{dR_L'}\,.
\end{equation}

One motivation to consider the analytic continued projected energy correlator $\frac{d\sigma^{[\nu]}}{d R_L}$ or cumulative distribution $\Sigma^{[\nu]}$
 is that, they have an approximate scaling behavior inside high-energy jets~\cite{Chen:2020vvp}, e.g. 
 \begin{equation}
 \Sigma^{[\nu]}(R_L) \sim R_L^{2\bar \gamma(\nu+1)}\,.
 \label{eq:sigma}
 \end{equation}
The exponent $\bar \gamma(\nu+1)$ can be expressed in terms of the DGLAP anomalous dimensions, specifically the $\nu$-moment of the splitting functions $P(x)$~\cite{Altarelli:1977zs,Gribov:1972ri,Dokshitzer:1977sg,Gross:1974cs,Gross:1973ju}. As a consequence of this, considering the $\nu$-correlator in the $\nu\to 0$ limit provides an alternative way to study small-$x$ physics~\cite{Lipatov:1996ts,Lipatov:1976zz,Kuraev:1976ge,Kuraev:1977fs,Balitsky:1978ic,Catani:1994sq,Ciafaloni:1998iv,Salam:1998tj,Ciafaloni:1999yw,Altarelli:1999vw,Altarelli:2001ji,Balitsky:1995ub,Kovchegov:1999yj,Jalilian-Marian:1996mkd,Jalilian-Marian:1997jhx,Iancu:2001ad}.
 The splitting function can also be exposed by the soft-drop momentum sharing variable~\cite{Larkoski:2015lea,Larkoski:2017bvj,Cal:2021fla}, but this explicitly excludes the small-$x$ region due to the $z_{\rm cut}$ from the soft drop condition.
The $\nu$-correlator with $\nu=0$ corresponds to the hadron multiplicity, whose scaling has been studied to high precision~\cite{Malaza:1985jd,Lupia:1997in,Eden:1998ig,Capella:1999ms,Bolzoni:2012ii,Bolzoni:2013rsa}.

 In perturbative QCD, the result of the anomalous dimension, $\bar{\gamma}(\nu+1)$, is analytic in spin $J=1+\nu$~\cite{Moch:2004pa,Vogt:2004mw,Chen:2020uvt,Gehrmann:2023iah,Moch:2021qrk,Falcioni:2023luc,Falcioni:2023vqq,Falcioni:2024xyt}, suggesting that it forms an analytic spectrum called a Regge trajectory.
 In the context of integrability or conformal symmetry, the concept of Regge trajectory and analyticity in spin are studied more rigorously in the past few decades~\cite{Gromov:2015wca,Caron-Huot:2017vep,Costa:2012cb,Homrich:2022mmd,Klabbers:2023zdz,Ekhammar:2024neh}. For example, in conformal theories, the current understanding is that while local operators are discrete in spin, light-ray operators are their analytically-continued version and form Regge trajectories~\cite{Kravchuk:2018htv,Balitsky:2013npa,Caron-Huot:2022eqs,Henriksson:2023cnh,Homrich:2024nwc}. 
Therefore, measuring the $\nu$-correlator in the high-energy jet ensembles provides an interesting way to  probe the analytic Regge trajectory as well as small-$x$ physics from the collider data.
However, the $\nu$-correlators have not been practically implemented for real events due to their highly complex definition, in contrast to efficient measurement of low integer $N$-point energy correlator~\cite{Komiske:2022enw, Budhraja:2024xiq}. Combining a recursion with the idea of \textsc{FastEEC}~\cite{Budhraja:2024xiq}, we here provide an efficient algorithm to measure the $\nu$-correlator and test it on CMS Open Data.

This letter is organized as follows: In Sec.~\ref{sec:nu}, we begin by reviewing the formal definition of the $\nu$-correlator. For completeness, we also briefly discuss the leading logarithmic (LL) result which exhibits an approximate scaling behaviour in the small angle limit. In Sec.~\ref{sec:method}, we discuss our recursive approach and its implementation in the \textsc{FastEEC} algorithm, that enables the computation of  $\nu$-correlators in a practical amount of time. We then describe the basic usage of our code~\cite{FASTEEC} in Sec.~\ref{sec:implementation}, available as an update to \textsc{FastEEC}. 
In Sec.~\ref{sec:EnuC}, we provide first results for non-integer values of $\nu$ computed on CMS 2011A Open Data~\cite{CMS:OpenAccessPolicy, CMS:JetPrimary2011A}. In Sec.~\ref{sec:results}, we utilize our method to extract the power law scaling (in the perturbative regime) and compare it against the analytic solution of the anomalous dimensions at LL. 
We conclude in Sec.~\ref{sec:conclusions}.  

\section{$\nu$-correlators}
\label{sec:nu}

In this section, we give a brief review of the $\nu$-correlator~\cite{Chen:2020vvp} and present a new recursion relation in \eqref{eq:nu-weights-compact} that will aid our implementation. 
For the integer $N$-point correlator in Eq.~\eqref{eq:PENC_def}, if we drop the measurement of the largest distance $R_L$ encoded in the delta function, we recognize the sum of products of momentum fractions as a multinomial expansion,
\begin{equation}
\sum_{i_1,i_2,\dots, i_N\in X}\!\!\! z_{i_1} z_{i_2} \dots z_{i_N}
\!=\!\biggl(\sum_{i\in X} z_i\biggr)^N 
\!\!\!=\! {\sum_{\alpha_i \geq 0}}^\prime
 \frac{\Gamma(N+1)}{\prod_{i\in X} \Gamma(\alpha_i+1)} \prod_{i\in X} z_i^{\alpha_i}
.
\end{equation}
Here $\sum^\prime$ means the sum is subject to the constraint $\sum_{i} \alpha_i = N$.
The multinomial expansion is broken by the $\delta$-function constraints on angular variables, which depends on the subset of particles $S\subset X$ whose energy weight contributes, 
\begin{equation}
\delta(R_L-\mathrm{max}\{\Delta R_{ij}\}_{i,j\in S})
 \biggl[{\sum_{\alpha_i \geq 1}}^\prime 
 \frac{\Gamma(N+1)}{\prod_{i\in S} \Gamma(\alpha_i+1)} \prod_{i\in S} z_i^{\alpha_i}
\biggr]\,.
\end{equation}
Note that the $\alpha_i$ here are only those for the subset $S$, hence $\alpha_i \geq 1$.
 To perform an analytic continuation from integer $N$ to non-integer $\nu$, we rewrite the constrained sum over exponents $\alpha_i$ as a sum over non-empty subsets $S^\prime \subset S$
\begin{equation}
 \biggl[{\sum_{\alpha_i \geq 1}}^\prime 
 \frac{\Gamma(N+1)}{\prod_{i\in S} \Gamma(\alpha_i+1)} \prod_{i\in S} z_i^{\alpha_i}
\biggr]= \sum_{S^\prime \subset S} (-1)^{|S|-|S^\prime|} \biggl(\sum_{z_i \in S^\prime} z_i\biggr)^N
\,.
\end{equation}
This naturally suggests how to analytically continue: simply replace $N$ with $\nu$ in the exponent 
\begin{equation}
\mathcal{W}^{[\nu]}(S)  = \sum_{S^\prime \subset S} (-1)^{|S|-|S^\prime|} \left(\sum_{z_i \in S^\prime} z_i\right)^\nu\,,
\label{eq:nu-weights-compact}
\end{equation}
such that the $\nu$-correlator is defined as
\begin{equation}\label{eq:PEnuC_def}
\frac{d\sigma^{[\nu]}}{d R_L} =\sum_X \int d\sigma_X \sum_{S\subset X} \mathcal{W}^{[\nu]}(S)\; \delta(R_L-\mathrm{max}\{\Delta R_{ij}\}_{i,j\in S})\,.
\end{equation}
Ref.~\cite{Chen:2020vvp} constructed $\mathcal{W}^{[\nu]}(S)$ in a recursive way
\begin{align}
&\mathcal{W}^{[\nu]}(\emptyset)=0\,, \nn \\
&\mathcal{W}^{[\nu]}(S) = \biggl(\sum_{i\in S} z_i \biggr)^\nu - \sum_{S^\prime \subsetneqq S} \mathcal{W}^{[\nu]}(S^\prime)\,.
\label{eq:nuweights}
\end{align}
The new recursion in \eqref{eq:nu-weights-compact}, introduced in this letter, will be important in speeding up the evaluation of $\nu$-correlators, as discussed in Sec.~\ref{sec:method}.

In the collinear limit $R_L\ll 1$, the $\nu$-correlator and its cumulative distribution $\Sigma^{[\nu]}$ can be factorized as a convolution of a hard function $\vec{H}$ and jet function $\vec{J}^{[\nu]}$~\cite{Dixon:2019uzg} 
\begin{equation}
\Sigma^{[\nu]}\Bigl(R_L, \ln \frac{p_{T,\text{jet}}}{\mu}\Bigr)\! =\!\! \int_0^1 \!\!\! \df x\, x^\nu \vec{J}^{[\nu]}\Bigl(\ln \frac{x p_{T,\text{jet}}R_L }{\mu}\Bigr)\cdot \vec{H}\Bigl(x, \ln \frac{p_{T,\text{jet}}}{\mu}\Bigr).
\end{equation}
Both the hard and jet functions are vectors in flavor space, i.e., $\vec{H} = \{H_q,H_g\}$ and $\vec{J} = \{J_q, J_g\}$ and each satisfies a RG equation with the time-like splitting function matrix $\hat{P}(y)$ as the evolution kernel
\begin{align}
&\frac{\df}{\df \ln \mu^2} \vec{H}\Bigl(x, \ln \frac{p_{T,\text{jet}}}{\mu}\Bigr) = - \int_x^1 \frac{\df y}{y} \hat{P}(y)\cdot \vec{H}\Bigl(\frac{x}{y},\ln \frac{p_{T,\text{jet}}}{\mu}\Bigr)\,, \nn\\
&\frac{\df}{\df \ln \mu^2} \vec{J}^{[\nu]}\Bigl( \ln \frac{p_{T,\text{jet}}R_L}{\mu}\Bigr) = \int_0^1 \df y \, y^\nu \vec{J}^{[\nu]}\Bigl(\ln \frac{y p_{T,\text{jet}}R_L}{\mu}\Bigr)\cdot \hat{P}(y)\,.
\end{align}
The evolution equations can be used to resum $\alpha_s^n \ln^m R_L$ series (with $m \leq n$) in $\Sigma^{[\nu]}$ that appear in the collinear limit $R_L \ll 1$. 

At leading-logarithmic accuracy, this leads to an elegant analytic result for $\Sigma^{[\nu]}$ 
\begin{equation} \label{eq:LL_result}
\Sigma^{[\nu]} = (1,1)\cdot \left(\frac{\alpha_s(p_{T,\text{jet}}R_L)}{\alpha_s(p_{T,\text{jet}})}\right)^{-\frac{\hat{\gamma}^{(0)}(\nu+1)}{\beta_0}}\cdot 
\begin{pmatrix}
f_q\\
f_g
\end{pmatrix}.
\end{equation}
Here $(f_q,f_g)$ provides the probability of quark and gluon jets for a given jet ensemble, $\hat{\gamma}^{(0)}(J)$ is the Mellin moment of the LO splitting kernel, 
$\hat{\gamma}(J) = - \int_0^1 dx\, x^{J-1} \hat{P}(x)$
which is also known as the DGLAP anomalous dimension for the spin-$J$ twist-2 light-ray operators. Explicitly, the anomalous dimension matrix is
\begin{equation}
\hat{\gamma}(J) =
\frac{\alpha_s}{4\pi} \begin{pmatrix}
\gamma^{(0)}_{qq}(J) & 2 n_f\gamma^{(0)}_{qg}(J)\\
\gamma^{(0)}_{gq}(J) & \gamma^{(0)}_{gg}(J)
\end{pmatrix} + \mathcal{O}(\alpha_s^2)\,,
\end{equation}
where
\begin{align}
\label{eq:gamma}
\gamma^{(0)}_{qq}(J) &= 4C_F \biggl[ \Psi(J+1)+\gamma_E- \frac{2}{J(J+1)} - \frac{3}{4}\biggr]\,, \\
\gamma^{(0)}_{gq}(J) &= -2C_F \frac{J^2+J+2}{(J-1)J(J+1)}\,,\nn\\
\gamma^{(0)}_{qg}(J) &=  -2T_F \frac{J^2+J+2}{J(J+1)(J+2)}\,,\nn \\
\gamma^{(0)}_{gg}(J) &= 4C_A\biggl[\Psi(J\!+\!1) \!+\! \gamma_E \!-\! \frac{1}{J(J\!-\!1)} \!-\! \frac{1}{(J\!+\!1)(J\!+\!2)}\biggr] - \beta_0\,.\nn 
\end{align}
Here $\Psi$ is the digamma function and $\beta_0=\frac{11}{3}C_A-\frac{4}{3}n_f T_F$ is the 1-loop QCD beta function constant.
Higher logarithmic resummation has been studied in~\cite{Chen:2020vvp,Lee:2022ige,Chen:2023zlx}. 

The consequence of Eq.~\eqref{eq:LL_result} is that $\Sigma^{[\nu]}$ exhibits an approximate power-law scaling in the small angle limit. Neglecting the matrix form in Eq.~\eqref{eq:LL_result}, corresponding to quark-gluon mixing, we find that 
\begin{equation}
R_L \frac{\df }{\df R_L}\, \Sigma^{[\nu]}(R_L) = 2 \bar{\gamma}(\nu+1)\, \Sigma^{[\nu]}(R_L)\,.
\end{equation}
In the presence of quark-gluon mixing, the exponent $\bar{\gamma}(\nu+1)$ should be interpreted as some effective mean of the two eigenvalues of $\hat \gamma$, where the relative contribution of the two eigenvalues also depends on ratio of quark and gluon jets.
Similarly, the differential distribution $\frac{d \sigma^{[\nu]}}{dR_L}$ also has an approximate scaling behavior with a shifted exponent $2[\bar{\gamma}(\nu+1) -\frac{\alpha_s}{4\pi}\beta_0]-1$. 

Based on the leading-logarithmic result, the scaling behavior of the $\nu$-correlator provides a way to access the continuous Regge trajectory in experiment. There are some subtleties: the scaling exponent also receives contributions from higher-loop corrections and the running coupling, and there are also hadronization effects.  Nevertheless, Eq.~\eqref{eq:gamma} gives a reasonable description of the $\nu$-correlator scaling exponent when $\nu \gtrsim 1$, as we will see in Sec.~\ref{sec:results}. 
However, as $\nu$ approaches $0$, this comparison breaks down due to the $1/\nu$ pole in the DGLAP  anomalous dimension. This limit is dual to the small-$x$ regime in the splitting kernel $\hat{P}(x)$ and is well-known to be governed by BFKL physics~\cite{Lipatov:1996ts,Lipatov:1976zz,Kuraev:1976ge,Kuraev:1977fs,Brower:2006ea,Kotikov:2000pm,Kotikov:2002ab}. As suggested in~\cite{Chen:2020vvp}, this is a potential way to look for BFKL physics in jets.

\section{Fast evaluation of $\nu$-correlators}
\label{sec:method}

For integer values of $N$, a direct implementation of the $N$-point correlators scales like $M^N/N!$, where $M$ is the number of particles in a jet. For the values of $N$ of interest in practice, $M \gg N$. In Ref.~\cite{Budhraja:2024xiq}, we used the idea that for correlations at a given angular scale the details at much smaller scales are irrelevant. One can, therefore, recluster the jet and calculate the energy correlator on subjets, yielding reliable results for angles sufficiently above the subjet radius. Since this approximation replaces the number of particles by the number of subjets, it yields a substantial speed up if the number of subjets is small. By employing a dynamical subjet radius, we obtained reliable results at all angular scales, finding speed ups of up to several orders of magnitude, depending on the desired level of accuracy.

The $\nu$-point correlator is more challenging. If the definition in Ref.~\cite{Chen:2020vvp} is implemented, here given in Eq.~\eqref{eq:nuweights}, this would scale as $2^{2M}$: it involves a sum over all subsets ($2^M$) of $M$ particles, and for each subset the contribution from all its subsets (another $2^M$) needs to be subracted. Therefore, only the $\nu$-correlator on final states with three particles was calculated in that paper, to verify the infrared safety of their analytic continuation. Even with our reclustering using a dynamic subjet radius, obtaining a reasonable level of precision within a practical time period is still prohibitive when using \eqref{eq:nuweights}.

To obtain results in a reasonable amount of time, we remove one $2^M$ by exploiting our recursion relation in \eqref{eq:nu-weights-compact}. We now show explicitly how this emerges from \eqref{eq:nuweights}, for a subset for $n$ particles, 
\begin{align}
 \mathcal{W}_n^{(\nu)}(i_1,\dots i_n) &= (z_{i_1} + ... + z_{i_n})^\nu \nn \\
	&\quad - \sum_{1\leq a_1<a_2< .. <a_{n-1}\leq n} \mathcal{W}_{n-1}^{(\nu)}(i_{a_1}, \dots i_{a_{n-1}}) \nn \\
	&\quad - \sum_{1\leq a_1<a_2< .. <a_{n-2}\leq n} \mathcal{W}_{n-2}^{(\nu)}(i_{a_1}, \dots i_{a_{n-2}}) \nn \\
	&\quad - \sum_{1\leq a_1<a_2< .. <a_{n-3}\leq n} \mathcal{W}_{n-3}^{(\nu)}(i_{a_1}, \dots i_{a_{n-3}}) \nn \\
	&\quad - \dots \nn \\
 &= (z_{i_1} + ... + z_{i_n})^\nu \nn \\
	&\quad - \sum_{1\leq a_1<a_2< .. <a_{n-1}\leq n} (z_{i_{a_1}} + z_{i_{a_2}} + ... +  z_{i_{a_{n-1}}})^\nu \nn \\
	&\quad + \sum_{1\leq a_1<a_2< .. <a_{n-2}\leq n} (z_{i_{a_1}} + z_{i_{a_2}} + ... + z_{i_{a_{n-2}}})^\nu \nn \\
	&\quad - \sum_{1\leq a_1<a_2< .. <a_{n-3}\leq n} (z_{i_{a_1}} + z_{i_{a_2}} + ... + z_{i_{a_{n-3}}})^\nu \nn \\
	&\quad + \dots
 \label{eq:weights}
\end{align}

We can thus calculate $\mathcal{W}_n^{(\nu)}$ recursively by subtracting the $\mathcal{W}_{n-1}^{(\nu)}$ for all its $n$ subsets of $n-1$ particles. For the terms involving $n-1$ particles this directly works. For the terms involving $n-2$ particles, we need to include a factor of $1/2!$, since there are two orders in which the two particles are removed, leading to double counting. Similarly, there is a $1/3!$ for the term involving $n-3$ particles. Since the term with $n-3$ particles in $\mathcal{W}_{n-1}$ already has a $1/2!$, we only need to multiply with $1/3$, etc. These extra factors are easily accommodated by keeping the terms with different numbers of particles separate in intermediate steps of the calculation. This leads to a substantial speed up of $2^{2M} \to M 2^M$, where $M$ appears from keeping the terms separate at each step of recursion. However, $M 2^M$ memory is now also required, so this is a trade-off between speed and memory. Furthermore, the largest distance of a subset of particles can be calculated recursively as well, otherwise that would require $M^2 2^M$ time. 

We implement this in an update to \textsc{FastEEC}. As in Ref.~\cite{Budhraja:2024xiq}, we recluster the jet, resolve subjets, calculate the energy correlator contribution on subjets involving both branches, and recurse on each branch. Concretely, we use the Cambridge/Aachen algorithm~\cite{Dokshitzer:1997in} to re-cluster the jet (with a large enough radius such that all particles remain in the reclustered jet). We then recursively traverse the clustering tree, at each step restricting to the contributions involving both branches of the splitting, for which the approximation to limit the angular resolution of the branches is justified. However,  to avoid memory problems, we don't use a fixed resolution parameter but a fixed maximum number of subjets instead. In our implementation we allow up to 8 subjets for each branch of a splitting, for a total of 16 subjets. 

\begin{figure}
		\centering
		\includegraphics[width=0.48\textwidth]{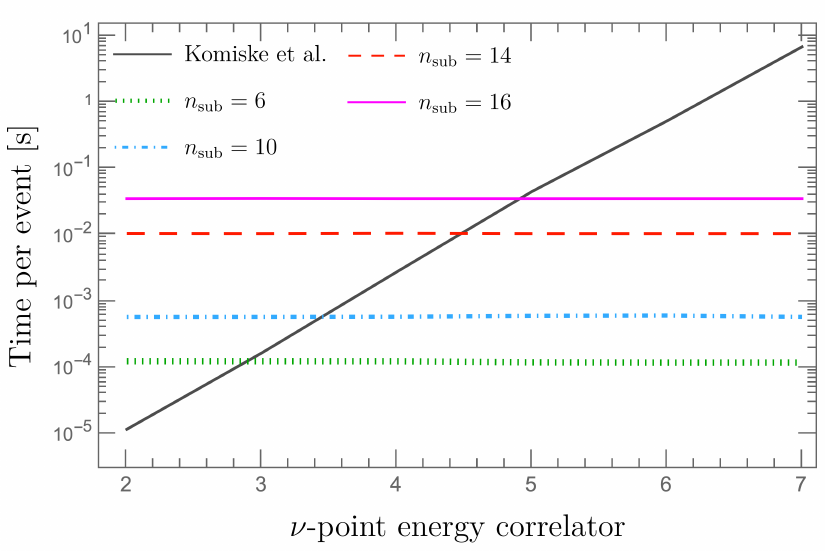}	
		\caption{Time per event for the case of integer $\nu$ values using our approach compared to Ref.~\cite{EEC_github}. The computation time of our method does not depend on the value of $\nu$ but on the number of resolved subjets, which determines the accuracy.}
		\label{fig:time}
\end{figure}

 \begin{figure*}[!htb]
	\centering 
	\includegraphics[width=0.495\textwidth]{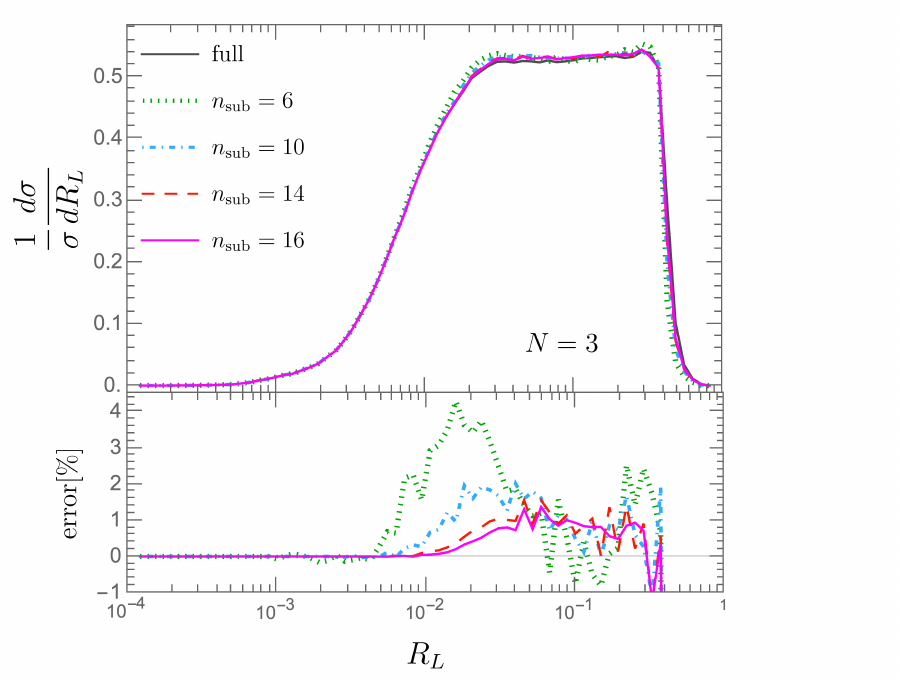}	
	\includegraphics[width=0.495\textwidth]{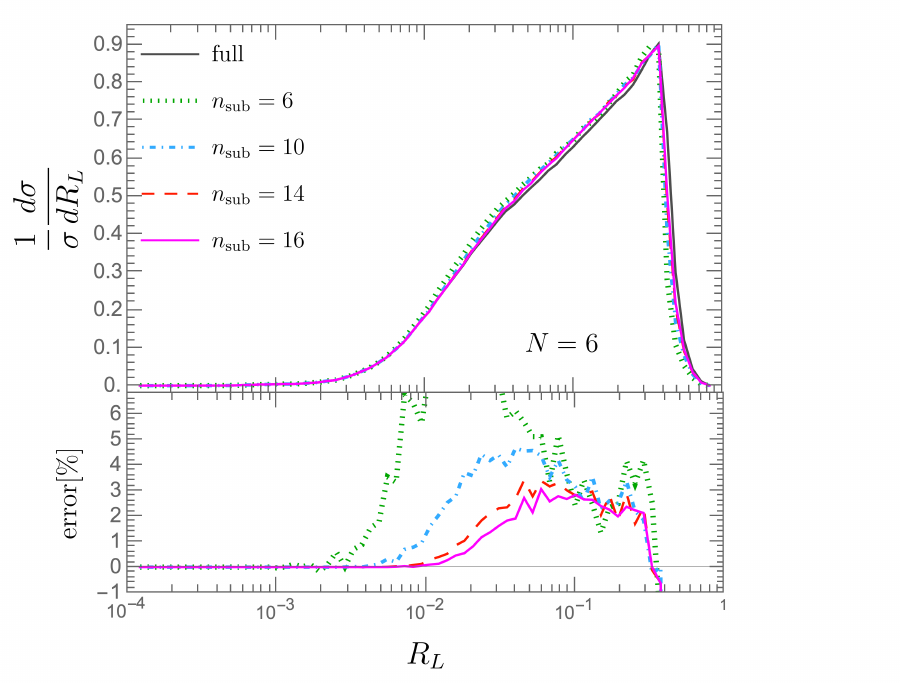}	\\
 \caption{$\nu$-point energy correlator for integer values: $N = 3$ (left) and $6$ (right), calculated using Ref.~\cite{EEC_github} and our approximate method with different maximum number of subjets. In the lower panel, the relative error of our method is shown, which decreases as $n_{\rm sub}$ increases. We find that for $n_{\rm sub} = 16$, the errors are only at the level of a (few) percent.}
	\label{fig:ENC}
\end{figure*}

Our approach makes it feasible to compute arbitrary $\nu$-correlators on realistic jets with a large number of particles. We demonstrate this by using the reprocessed data on jets from the CMS 2011A Open Data~\cite{CMS:OpenAccessPolicy,CMS:JetPrimary2011A}. In Fig.~\ref{fig:time}, we show the time per event (in seconds) for the case of (integer) $\nu$ values with different maximum number of subjets $n_{\rm sub}$, compared to the standard computation of energy correlators for integer $N$ implemented through the \verb+EnergyEnergyCorrelators+ package~\cite{EEC_github}. These results were obtained with a single core on a M2 MacBook Air. We see that for $n_{\rm sub} = 16$ the amount of time needed is roughly ${\cal{O}}(0.01 s)$ per jet and, as we will show later, this corresponds to about a percent-level precision of the energy correlator distribution.

We note that our implementation of the $\nu$-correlator is independent of the specific value of $\nu$, unlike our previous implementation in Ref.~\cite{Budhraja:2024xiq}, where the computation time grows with the value of (integer) $\nu$. However, the point at which these implementations require a similar amount of time, i.e.~$M^N/N! \sim M2^M$ corresponds to $M \sim N$. Since $M \gg N$ in practice, our previous implementation is faster for integer $N$-point correlators.\footnote{This may seem at odds with the (slow) exponential increase in computation time with $N$ of the original implementation in \textsc{FastEEC}, whereas here the computation time is independent of $N$. However, here we fix the maximum number of subjets whereas we previously had a fixed resolution parameter, which is responsible for this difference.}

\section{Implementation and consistency checks}
\label{sec:implementation}

\begin{figure*}[!htb]
	\centering 
	\includegraphics[width=0.48\textwidth]{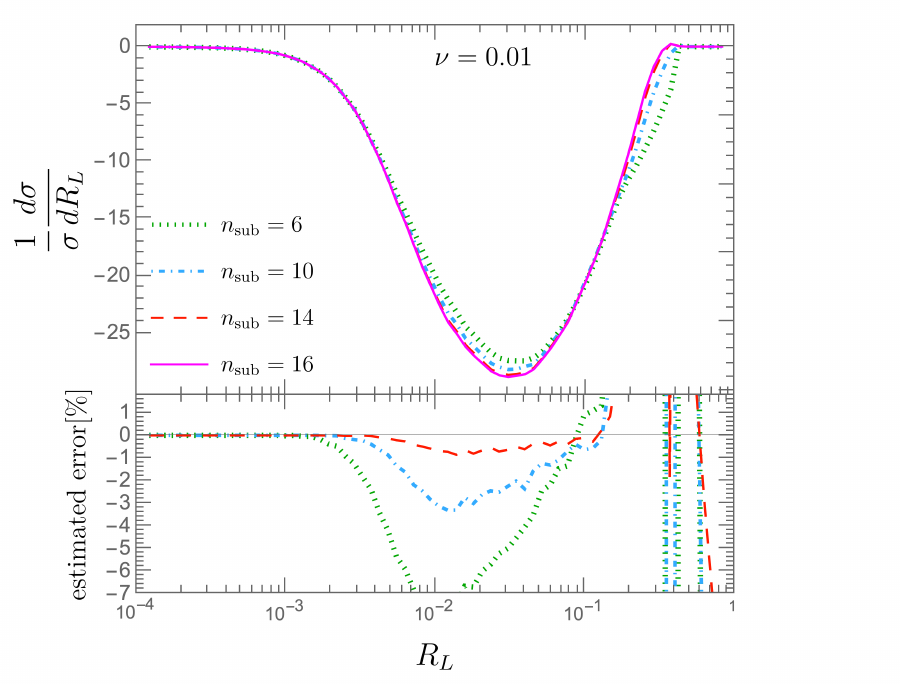}	
	\includegraphics[width=0.48\textwidth]{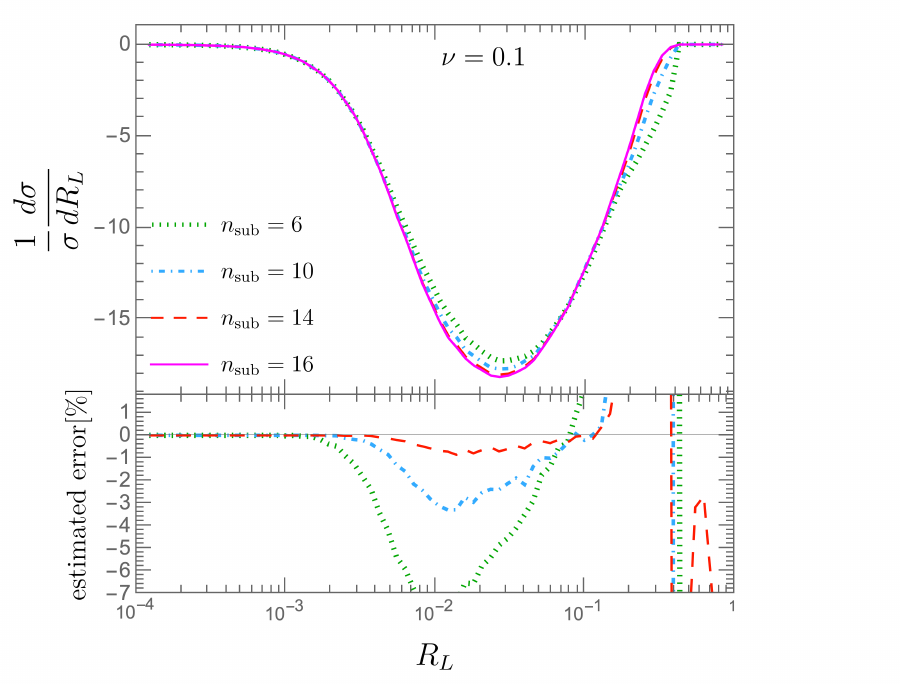}	\\
    \hspace{0.2cm}\includegraphics[width=0.48\textwidth]{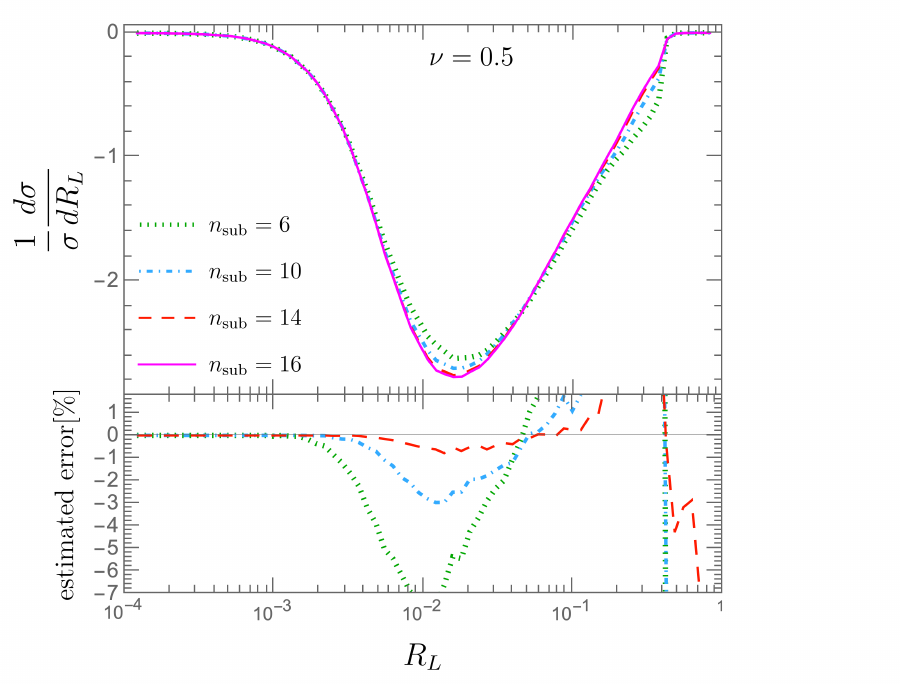}	\includegraphics[width=0.48\textwidth]{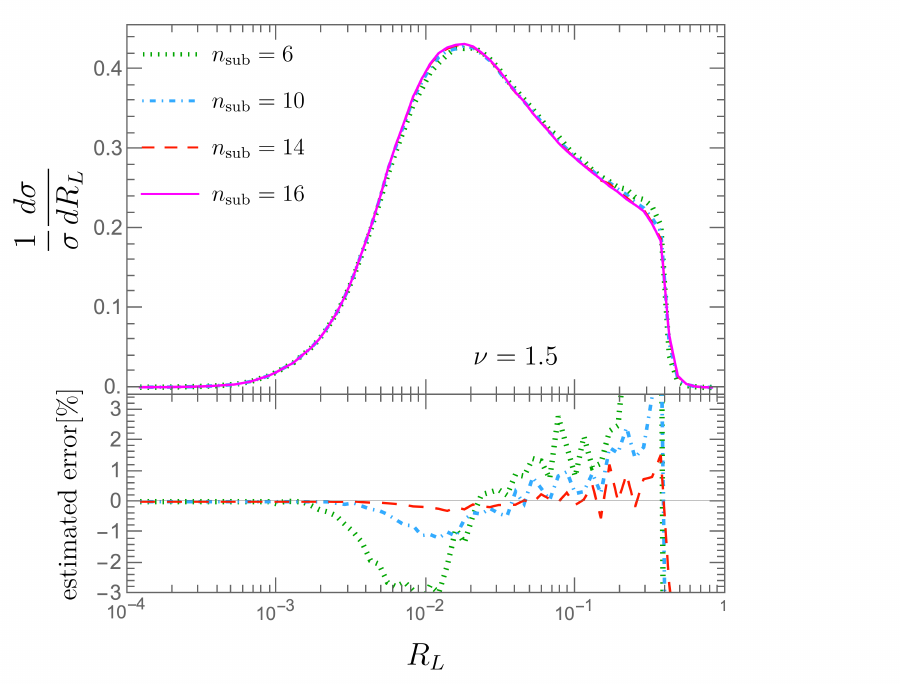}	\\
 \caption{$\nu$-point energy correlator distribution for non-integer $\nu$ values. The different panels correspond to $\nu= 0.01, 0.1,0.5$ and $1.5$ respectively, and the $\nu$-correlator approximated with a different maximum number of subjets is shown. In the lower panel, we present an error estimate by comparing to our best result, $n_{\rm sub}=16$. We find that, for $R_L \lesssim 0.1$, the differences are less than a percent between $n_{sub}=14$ and $n_{\rm sub}=16$, indicating a reasonable convergence has been achieved.}
	\label{fig:convergence}%
\end{figure*}

Our code for $\nu$-point energy correlators is released as an update~\cite{FASTEEC} of the \textsc{FastEEC} code. The code is provided in the file \verb+eec_fast_nu_point.cc+. 
 The command line inputs remain the same as with \textsc{FastEEC} except that the jet resolution parameter is replaced by the maximum number of subjets $n_{sub}$.  Explicitly, the command line inputs read as
\begin{align*}
\texttt{
./eec\_fast\_nu\_point input\_file events $\nu$ $n_{\rm sub}$} \\
\texttt{minbin nbins output\_file}
\end{align*}
Here \verb+input_file+ specifies the dataset from which events should be read. The \verb+events+ is the total number of jet events that need to be analysed, $\nu$ specifies which $\nu$-correlator to compute (called $N$ internally in the code to match the previous version), $n_{\rm sub}$ is the maximum number of subjets that should be used, \verb+minbin+ is the logarithm of the minimum bin value in the histogram, \verb+nbins+ is the total number of bins in the histogram and \verb+output_file+ is the name of the output file in which the result needs to be stored. The \verb+maxbin+ value is set to 0 corresponding to $R_L = 1$. The histogram that is output is normalized, with the lowest and highest bin corresponding to the underflow and overflow bins, respectively.
We provide reprocessed CMS Open Data on jets converted into a text format that can be used as the input set. 

We implement the recursive computation of the weights as outlined in Eq.~\eqref{eq:weights}, finding the largest separation for the given subset of particles, at each step of the recursion. 
We limit the number of subjets $n_{\rm sub}$ that can be used to a maximum value of 16, but it is straightforward to extend this to higher values, if needed. 
We have explicitly verified that our implementation satisfies the sum rule
\begin{equation}
    \int_0^1\! \df R_L\, \frac{\df \sigma^{[\nu]}}{\df R_L} = \sigma_{\rm tot} 
\end{equation}
for all values of $\nu$.

We now validate our new code by comparing to the known results for integer $N$, obtained using the \verb+EnergyEnergyCorrelators+ package~\cite{EEC_github}. This comparison is shown in Fig.~\ref{fig:ENC} for two different values, $N=3$ and $6$. In the lower panel, we also display the relative error between our approximate method and the exact computation. We see that the error decreases as we increase the maximum number of resolved subjets $n_{\rm sub}$. For $n_{\rm sub} = 16$, the relative error in the perturbative regime is about $\sim 1\%$ for $N=3$ and $\sim 2-3 \%$ for $N=6$. The different approximations disagree near the jet boundary where the use of subjets breaks down and the detailed position of particles becomes important. However, most phenomenological applications focus on the region away from the jet boundary, since the behavior of the energy correlator in this regime is largely determined by details of the choice of jet clustering algorithms rather than physics.

\section{Numerical results for non-integer $\nu$ values}
\label{sec:EnuC}

\begin{figure*}[!htb]
	\centering 
	\includegraphics[width=0.48\textwidth]{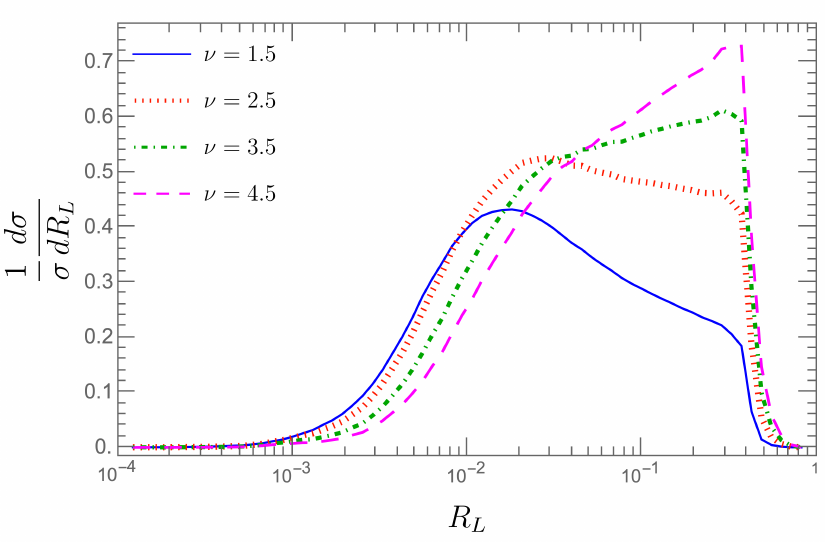}	
	\includegraphics[width=0.48\textwidth]{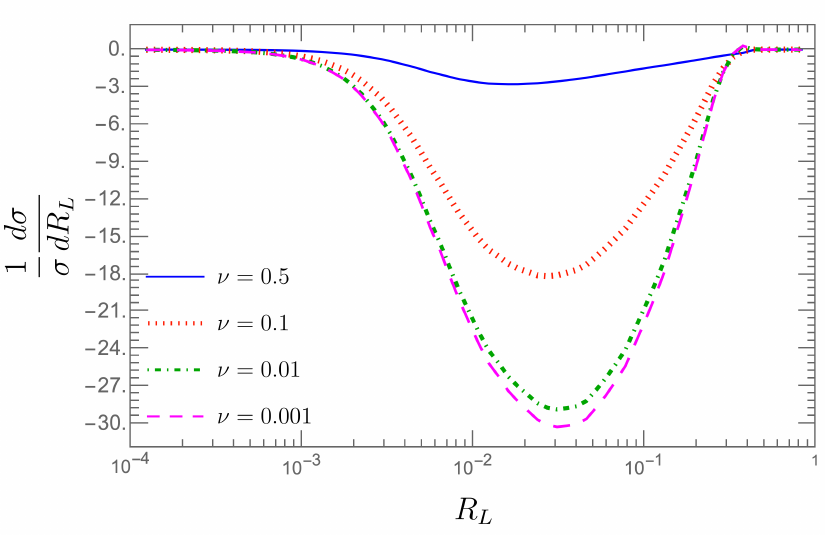}	\\
 \caption{$\nu$-point energy correlator for non-integer $\nu$ values for $\nu > 1$ (left) and $\nu<1$ (right) when using a maximum of 16 subjets. For $\nu > 1$, we see that the transition region from perturbative scaling regime to the non-perturbative regime shifts towards the right as $\nu$ increases. 
 For $\nu < 1$, the transition region again shifts to the right as one approaches $\nu$ close to 0.
 We also observe that as $\nu$ approaches 0, the $\nu$-correlator saturates.}
	\label{fig:EnuC}%
\end{figure*}

In this section, we provide first results for non-integer $\nu$-correlators. The results presented here are obtained from the CMS 2011A Open Data, available in a publicly usable format through the MIT Open Data initiative~\cite{Komiske:2019jim, MITOpenData, CMS:JetPrimary2011A, CMS:OpenAccessPolicy}. The dataset contains jets with a transverse momenta in the range $p_T \in [500, 550]$ GeV and with pseudo-rapidity $\vert \eta \vert < 1.9$. 

For non-integer values of $\nu$, we start by checking the convergence of our method as the maximum number of subjets is increased. We show this in Fig.~\ref{fig:convergence} for various $\nu$ values, i.e., $\nu = 0.01, 0.1, 0.5$ and $1.5$, finding a good convergence to the result with a maximum of 16 subjets, away from the jet boundary. 
In the lower panel, we estimate the error of our method by comparing $n_{\rm sub} = 6,10,14$  to the $n_{\rm sub}=16$ case, finding that there is less than a percent  difference  between the curves for $n_{\rm sub} =14$ and $n_{\rm sub}=16$ for $R_L\lesssim 0.1$ and all values of $\nu$ shown here. 

Next, in Fig.~\ref{fig:EnuC}, we compare the differential distributions of the $\nu$-correlators for various non-integer $\nu$ values. 
The $\nu$-correlator exhibits a different behaviour depending on whether $\nu > 1$ or $\nu<1$. 
While for $\nu > 1$, the distributions are positive, they become negative when $\nu <1$ and collapse to a delta function $\delta(R_L)$ for $\nu=1$.
Additionally, we note that for $\nu < 1$, the peak of the distribution becomes more and more pronounced as $\nu$ approaches 0, saturating at very small values of $\nu$. 
From the differential distributions we also observe that for $\nu>1$, the transition between the perturbative and non-perturbative regions, which have distinct power-law scaling, shifts to the right (larger $R_L$) for increasing values of $\nu$. Interestingly, this transition also shifts to the right when $\nu$ approaches 0. 

Following the arguments of Ref.~\cite{Lee:2024esz} for the leading non-perturbative contribution, we find that the contribution to the $\nu$-correlator for a soft gluon emission is notably different depending on whether $\nu>1$ or $\nu<1$. In particular,

\begin{equation}
\label{eq:nonp}
(z_c+z_s)^{\nu} - z_c^{\nu} - z_s^{\nu} = 
\begin{cases}
  \, \nu \, z_s \, z_c^{\nu-1} &\quad {\text{for} }\,\, \nu \gg 1\, , \\
  \,  -z_s^{\nu} &\quad {\text{for} }\,\, \nu\ll 1\, .
\end{cases} 
\end{equation}
where $z_s$ is the transverse momentum fraction of the soft gluon and $z_c$ that of an energetic particle. 
In Ref.~\cite{Lee:2024esz} it was shown that for integer values of $\nu>2$ this implies that the peak position, corresponding to the transition point, can be approximately described by $R_L^{[\nu],\rm peak} \sim (\nu/2)\, R_L^{[\nu=2],\rm peak}$ relative to the two-point energy correlator. This result can be analytically continued for $\nu>1$.

On the other hand, for $\nu<1$, Eq.~\eqref{eq:nonp} implies that $\nu$-dependence of the transition point should appear in the power-law exponent.  From the CMS open data analysis we find that the transition point has a weak dependence on $\nu$, and further investigation is needed to appropriately understand the size of non-perturbative contributions in the region for $\nu<1$. We leave this for a future study.

While the differential distributions make it easier to read off the transition between the non-perturbative (free hadron) and perturbative (quarks and gluons) region due to their different scaling behavior, the scaling behaviour of the $\nu$-correlators as function of $\nu$ is more apparent in the cumulative distribution in Fig.~\ref{fig:cumulant}.
From the figure, we clearly see that the slope, in the perturbative region, is positive for $\nu>1$, vanishes when $\nu=1$ and becomes negative for $\nu<1$. We will utilize the cumulative distribution to extract this power-law for a wide range of $\nu$ values in the next section.

\section{Extracting the anomalous dimensions}
\label{sec:results}

To showcase the \textsc{FastEEC} algorithm for the $\nu$-correlator, we extract its scaling exponent in jets using CMS Open Data, and briefly comment on the physics interpretation.
We do not provide a high-precision analysis but rather aim to motivate more detailed experimental measurements that  address uncertainties, as well as more precise theoretical calculations beyond our simple leading-logarithmic analysis.

\begin{figure}[t]
	\centering 
	\includegraphics[width=1\linewidth]{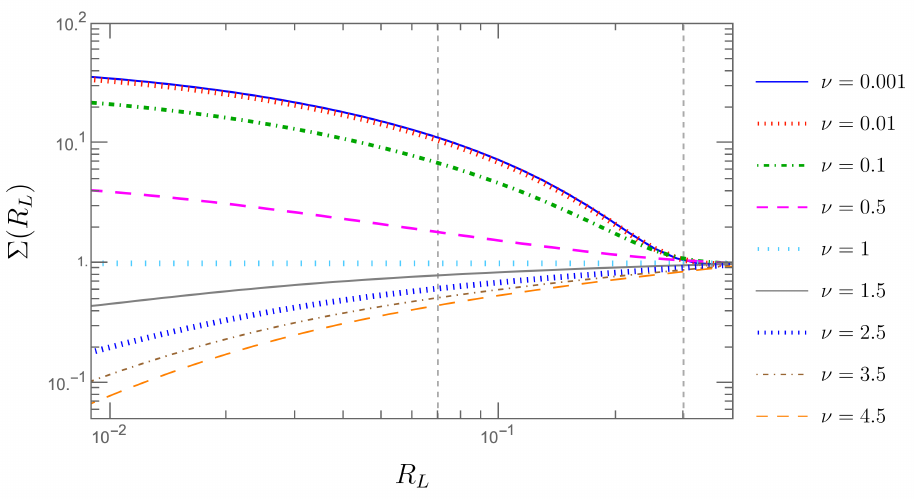}	
 \caption{Cumulative distribution of $\nu$-point energy correlator for various non-integer $\nu$ values. While the slope is positive for $\nu>1$, it changes sign for $\nu<1$ and vanishes when $\nu=1$. The vertical dashed lines represent the fit region used to extract the power law scaling from the Open Data in Sec.~\ref{sec:results}. }
	\label{fig:cumulant}
\end{figure}

We prefer to use the cumulative $\nu$-correlator to extract the scaling exponent for its smoother behavior in $\nu$, particularly near $\nu=1$, compared to its differential counterpart. Based on the differential distributions in Fig.~\ref{fig:EnuC} and the discussion of the transition to the non-perturbative region in Sec.~\ref{sec:EnuC}, we choose the fit region $0.07<R_L<0.3$ to avoid both the (transition to the) non-perturbative region and jet boundary effects, indicated with vertical dashed lines in Fig.~\ref{fig:cumulant}.  The errors shown are the 2$\sigma$ fit errors. 
We have also explored varying the fit range by 10\%, finding that this is not more than the fit error.

\begin{figure}
    \centering
    \includegraphics[width=1\linewidth]{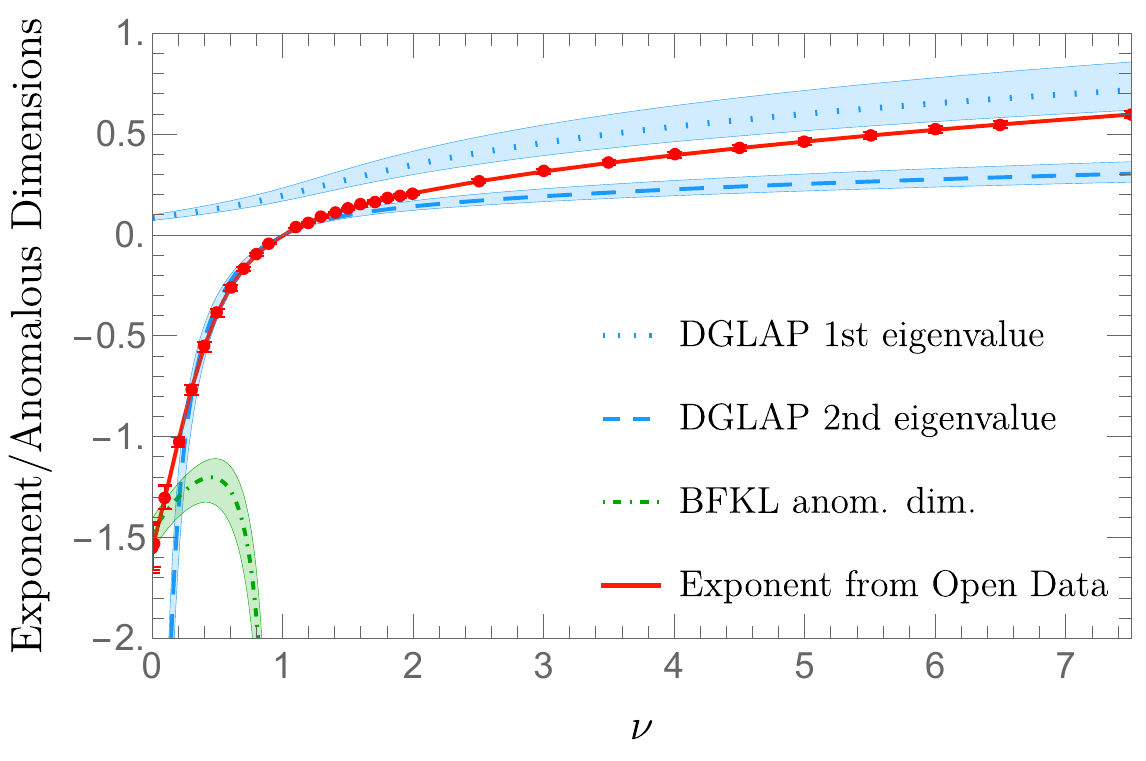}
    \caption{
    Comparison between the $\nu$-correlator scaling exponent and DGLAP/BFKL anomalous dimensions. Red dots are fitted exponents for $\nu$-correlator with the $2\sigma$ fit error. The light-blue bands and light-green band are the corresponding eigenvalue(s) of DGLAP anomalous dimension and BFKL anomalous dimension, respectively. Both bands are plotted by varying scale in $\alpha_s$ from $\mu^*/2$ to $2\mu^*$, with the central dashed line  evaluated at $\alpha_s(\mu^*) = 0.178$.}
    \label{fig:exponent_plot}
\end{figure}

As stated in Sec.~\ref{sec:nu}, the scaling exponent of cumulative $\nu$-correlator is proportional to anomalous dimension. However, due to the existence of quark-gluon mixing, the anomalous dimension is a $2\times 2$ matrix, whose eigenvalues should be averaged in a way that depends on the quark/gluon jet ratio. Indeed, Fig.~\ref{fig:cumulant} indicates that a fit to a single power-law, though good for most values of $\nu$, does not work as well for small $\nu$, leading to larger uncertainties in Fig~\ref{fig:exponent_plot}.
Without estimating the quark/gluon jet ratios, we simply compare our extracted exponent to both of these eigenvalues, from which we can still get reasonable information. 

The anomalous dimension depends on strong coupling constant $\alpha_s$ and hence is a scale-dependent quantity. The physical scale in this problem should be chosen as the jet scale, which is proportional to $p_T R_L$. There is no rigorous first-principle calculation known yet to determine the exact effective scale. Based on the experience from Ref.~\cite{Dixon:2019uzg}, for $p_T\sim 500\, \mathrm{GeV}$ jet with $\nu$-correlator angle $R_L\sim 0.1$, we believe an effective scale $\mu^*= 10\, \mathrm{GeV}$ is an appropriate approximation. We take $\alpha_s(m_Z)=0.118$ and use the two-loop running coupling, resulting in $\alpha_s(\mu^*) = 0.178$. The comparison of the extracted exponent to the eigenvalues of the leading-order anomalous dimension is shown in Fig.~\ref{fig:exponent_plot}.

For $\nu \gtrsim 1$, the DGLAP anomalous dimension gives a reasonable estimate for the scaling exponent of $\nu$-correlator, 
which is illustrated in Fig.~\ref{fig:exponent_plot} by the extracted scaling exponent lying between two eigenvalues of DGLAP anomalous dimensions.
When $\nu$ is close to $0$, one of the DGLAP anomalous dimensions in $\hat{\gamma}(\nu+1)$ eigenvalues diverges and no longer matches the $\nu$-correlator exponent. Instead, we expect the scaling exponent can be approximated by $-1+\nu+\gamma_{\text{BFKL}}(\nu)$~\footnote{We thank Cyuan-Han Chang, Petr Kravchuk, David Simmons-Duffin and HuaXing Zhu for private discussions on related questions.}, where $\gamma_{\text{BFKL}}(\nu)$ is the well-known BFKL eigenvalue
\begin{equation}
\gamma_{\text{BFKL}}(\nu) = \frac{\alpha_s}{\pi} C_A \biggl[
\Psi\biggl(\frac{\nu+1}{2}\biggr)+\Psi\biggl(\frac{-\nu+1}{2}\biggr)+2\gamma_E
\biggr]+\mathcal{O}(\alpha_s^2)\,.
\end{equation}
We find that this BFKL eigenvalue provides a reasonable prediction for the intercept value at $\nu=0$, as shown in Fig.~\ref{fig:exponent_plot}. Further details and a more thorough analysis is left to future work.

\section{Conclusions}
\label{sec:conclusions}

In this letter, we developed a method for evaluating $\nu$-correlators, the analytic continuation of the projected (integer) $N$-point energy correlators, that is sufficiently fast to be used on real data. The $\nu$-correlator provides access to non-integer moments of the splitting functions. A better understanding of the $\nu$-correlator over the whole $\nu>0$ range also provides valuable information for Monte Carlo parton showers. The limit $\nu \to 0$ is particularly interesting,  providing a way to probe small-$x$ physics in jets! 

In the original formulation, the computation of the $\nu$-correlator for a jet with $M$ particles involves a recursion over all $2^M$ subsets, and for each of these subsets the contribution from all other subsets need to be subtracted, yielding a $2^{2M}$ scaling. Here we trade memory for speed, using a new recursion relation to reduce the computation time to $M 2^M$, at the expense of needing order $M 2^M$ memory. 
 
We implement this in \textsc{FastEEC}, replacing the number of particles $M$ in the jet with the number of dynamically resolved subjets, capped to a maximum value $n_{\rm sub}$. In practice, we find that using $n_{\rm sub}= 16$ is sufficient to obtain results with precision at the (few) percent level, in the perturbative power-law region. Utilizing our efficient approach, we present initial results for the $\nu$-correlator using CMS Open Data. 

As a first foray into phenomenology, we use this to extract the power-law scaling in the perturbative region and compare it with the anomalous dimensions at leading logarithmic accuracy. We find that the extracted power law from the Open Data is in reasonable agreement to the DGLAP evolution except for $\nu \ll 1$. Here the DGLAP anomalous dimension is known to exhibit a $1/\nu$ divergence requiring resummation. Instead, we find that the extracted value as $\nu \to 0$ is well approximated by the corresponding eigenvalue of the BKFL anomalous dimension.

We hope that our analysis strongly motivates measurements of $\nu$-correlators on jets, that account for uncertainties, as well as high-precision theoretical studies, beyond the simple leading logarithmic accuracy that we restrict to and including the effect of hadronization. We emphasize that our approach makes it practical to put these theoretical ideas to the test on real-world LHC jets.

\section*{Acknowledgements}
We thank C.H. Chang, P. Kravchuk, K.~Lee, I.~Moult, D. Simmons-Duffin and H.X.~Zhu for discussions, and I.~Moult and H.X.~Zhu for feedback on this manuscript. A.B.~is supported by the project 'Microscopy of the Quark Gluon Plasma using high-energy probes' (project number VI.C.182.054) which is partly financed by the Dutch Research Council (NWO). H.C.~is supported by the U.S. Department of Energy, Office of Science, Office of Nuclear Physics under grant Contract Number DE-SC0011090. 
H.C. and W.W. would like to express special thanks to the Mainz Institute for Theoretical Physics (MITP) of the Cluster of Excellence PRISMA+ (Project ID 390831469) for its hospitality and support.

\bibliographystyle{elsarticle-num}
\bibliography{EEC_ref}

\end{document}